\documentclass[twocolumn,           
               showpacs,            
               nopreprintnumbers,   
               aps,                 
               prd,          	    
               a4paper,             
               groupedaddress,      
               groupedaddress,  
               unsortedaddress,
               nofootinbib,         
               tightenlines,        
               floats               
               ]{revtex4}

\usepackage{graphicx}
\usepackage{subfigure}
\usepackage{latexsym}
\usepackage{amssymb,latexsym}
\usepackage[draft=false]{hyperref}

\begin{document}




\title{Nflation: observable predictions from the random matrix mass
  spectrum} 
\author{Soo A Kim and Andrew R.~Liddle}
\affiliation{Astronomy Centre, University of Sussex, Brighton BN1 9QH,
  United Kingdom}
\date{\today} 
\pacs{98.80.Cq \hfill astro-ph/yymmnnn}
\preprint{astro-ph/yymmnnn}


\begin{abstract}
We carry out numerical investigations of the perturbations in Nflation
models where the mass spectrum is generated by random matrix theory.
The tensor-to-scalar ratio and non-gaussianity are already known to
take the single-field values, and so the density perturbation spectral
index is the main parameter of interest. We study several types of
random field initial conditions, and compute the spectral index as a
function of mass spectrum parameters. Comparison with microwave
anisotropy data from the Wilkinson Microwave Anisotropy Probe shows
that the model is currently viable in the majority of its parameter
space.
\end{abstract}

\maketitle

\section{Introduction}

The Nflation model of Dimopoulos et al.~\cite{DKMW} corresponds to a
collection of uncoupled massive fields which drive inflation via the
assisted inflation mechanism \cite{LMS}. The existence of multiple
fields is motivated by the axions of string theory, and their presence
enables sufficient inflation to be obtained without using
super-Planckian field values. Models of this type had first been
considered by Kanti and Olive \cite{KO} and then Kaloper and Liddle
\cite{KL} in the context of Kaluza--Klein models. They showed the
massive fields evolve faster to the minima of their own potential and
light ones later. Easther and McAllister~\cite{EM} introduced
random matrix theory as a way of computing the possible distribution
for masses in the Nflation model. For some related constructions see
Ref.~\cite{mfield}. 

It is obviously important to develop observational predictions from
such models. This has been thought difficult, because in multi-field
models the predictions depend in general upon the field initial
conditions as well as the model parameters. Easther and McAllister
\cite{EM} only studied two types of initial conditions, where either
the field values or the field energy densities were equal. Neither is
well motivated physically. However, in the case of an exponential mass
spectrum, Kim and Liddle \cite{KL1} showed that provided there were
enough fields, with randomly-chosen initial conditions, the
observational predictions become essentially \emph{independent} of
initial conditions again. The reason is that with enough fields, the
space of possible initial conditions is well sampled by a single
realization, and they named this the `thermodynamic regime'.

The purpose of this article is to apply the random initial conditions
approach of Ref.~\cite{KL1} to the random matrix mass spectrum of
Ref.~\cite{EM}, in order to explore its observational predictions and
test its viability.

The main observables are the density perturbation spectral index
$n_{{\rm S}}$, the tensor-to-scalar ratio $r$, and the non-gaussianity
parameter $f_{\rm NL}$. Some quite general results are already known,
applying to arbitrary mass spectra and initial conditions provided
sufficient inflation is obtained. Alabidi and Lyth \cite{AL} showed
that $r$ always has the same value as in the single-field case, and
Kim and Liddle \cite{KL2} showed that the same was true of $f_{\rm
NL}$ (this having previously been shown for two fields in
Ref.~\cite{VW}). Only $n_{{\rm S}}$ has model and initial conditions
dependence, and it has been shown that its predicted value cannot be
larger than the single-field value \cite{LR,piao}. There are some
further generalisations of these results \cite{piao,KL2}.

\section{The random matrix theory}

Once one considers the scalar fields in inflation as axions in string
theory, their masses can be written in matrix form, which depends on
specific details on compactification (following Ref.~\cite{EM}, we
also assume that higher-order terms can be neglected, so that what are
really cosine functions can be approximated as massive uncoupled
fields). The shape of the mass distribution depends only on the basic
structure of the mass matrix, which is specified by the supergravity
potential. In the simplest assumption, the entries in the mass
spectrum are independent and identically distributed, i.e.\ a random
matrix. The fields can be uncoupled by diagonalization of this matrix,
with the mass spectrum given by the distribution of eigenvalues.  The
distribution of the eigenvalues for random matrices of this kind is
characterized by the Mar\u{c}enko--Pastur law \cite{MP} when the
matrices are large. The distribution function depends on a parameter
$\beta$, the ratio of the number of axions to the dimension of the
moduli space. Easther and McAllister \cite{EM} devised this formalism
and computed the observational predictions in terms of $\beta$ for
specific choices of initial conditions for the fields where the field
values or the energy densities are identical (see also Ref.~\cite{gong}).

\begin{figure}[t]
\centering
\includegraphics[width=7.5cm]{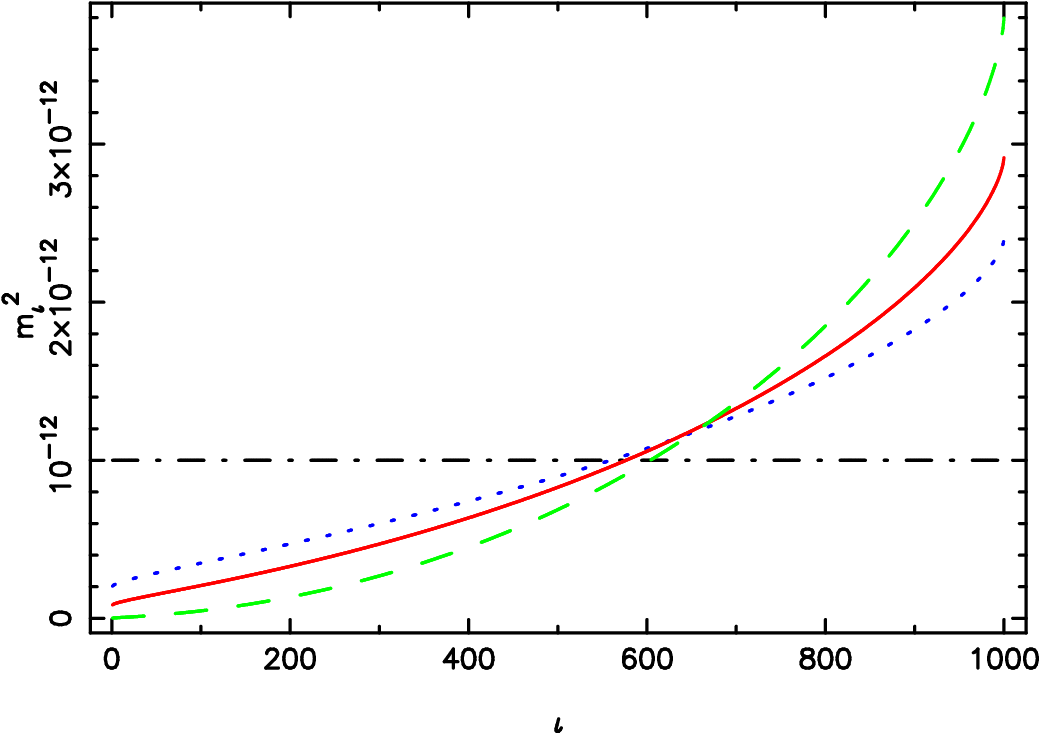}
\caption{\label{f:mbeta} The mass-squared distribution for 1000
  fields; the dashed-dotted (black) flat line is for $\beta=0$, the dotted
  (blue) for $\beta=0.3$, the solid (red) for $\beta=0.5$ and the
  dashed (green) for $\beta=0.9$.}
\end{figure}

We follow their notation for the mass spectrum, and label the average
value of the mass-squared $\langle m^2 \rangle =\bar{m}^2$. We will
throughout put $\bar{m}=10^{-6}\, M_{\rm Pl}$ where $M_{\rm Pl}$ is
the reduced Planck mass; a simple rescaling of $\bar{m}$ would be
sufficient to match the observed normalization of perturbations.  The
mass spectrum is determined by two quantities, the number of fields
$N_{\rm f}$ and a parameter $\beta$ which governs its shape. The shape
parameter $\beta$ lies in the range zero to one, with values around
$1/2$ perhaps the most plausible \cite{DKMW,EM}. Figure~\ref{f:mbeta}
shows the shape of the spectrum for $N_{\rm f} = 1000$ and some values
of $\beta$.

\section{Observational predictions}

For a set of uncoupled fields with quadratic potentials, the number of
$e$-foldings $N$, in the slow-roll approximation, is~\cite{LR}
\begin{equation}
N \simeq \frac{\sum_i \phi_i^2}{4M_{\rm Pl}^2},
\label{e:efold}
\end{equation}
where $\phi_i$ is the $i$-th field. For our initial results, we choose
the field initial values randomly from a uniform distribution in the
range $0$ to $M_{{\rm Pl}}$.  The total number of $e$-foldings is then
accurately given by the linear relation $N_{{\rm tot}} \simeq N_{\rm
f}/12$ \cite{KL1}, this result holding for an arbitrary mass
spectrum. We will throughout assume the observable scales crossed
outside the horizon 50 $e$-foldings before the end of inflation.
Accordingly, sufficient inflation requires a minimum of around 700
fields.

In passing we note that increasing the number of fields increases the
energy scale at the end of inflation, and hence might increase the number
of $e$-foldings relevant to observable perturbations. However this
effect is well within current uncertainties from the unknown behaviour
of the Universe between inflation and nucleosynthesis.

We follow the usual formulae for the observational predictions
of ${\cal P}_{\cal R}$, $n_{\rm S}$, $r$, and $f_{\rm NL}$.
These are \cite{LR,EM,KL1,KL2}
\begin{eqnarray}
{\cal P}_{\cal R} &\simeq& \frac{\sum_i m_i^2 \phi_i^2 \sum_j
\phi_j^2}{96 \pi^2 M_{\rm Pl}^6}\,,
\label{e:P}\\
n_{\rm S} &\simeq& 1 - 4 M_{\rm Pl}^2 \left[ \frac{\sum_i m_i^4
\phi_i^2}{(\sum_k m_k^2 \phi_k^2)^2} + \frac{1}{\sum_j
\phi_j^2}\right]\,,
\label{e:ns}\\
&\simeq& 1 - \frac{1}{N}-4 M_{\rm Pl}^2 \left[ \frac{\sum_i m_i^4
\phi_i^2}{(\sum_k m_k^2 \phi_k^2)^2}\right]\,, 
\label{e:ns1}\\
r &\simeq& \frac{32 M_{\rm Pl}^2}{\sum_i \phi_i^2} \simeq
\frac{8}{N}\,,\label{e:r}\\ -\frac{5}{6}f_{\rm NL}^{(4)} &\simeq&
\frac{2 M_{\rm Pl}^2}{\sum_i \phi_i^2} \simeq \frac{1}{2N} \simeq
\frac{r}{16}\,,
\label{e:fnl}
\end{eqnarray}
where $m_i$ and $f_{\rm NL}^{(4)}$ are the $i$-th mass and the second
term of the nonlinearity parameter $f_{\rm NL}$ respectively (the
first contribution to $f_{\rm NL}$ is model-independent and small
\cite{Maldacena, wmap3}). More
detailed calculations of the non-gaussianity have confirmed that this
is the leading contribution to the non-gaussianity \cite{Ebat}.  We
have simulated these observables and will show in particular $n_{\rm
S}$ to study the effect from this mass distribution.

\subsection{The spectral index}

\begin{figure}[t]
\centering
\includegraphics[width=7.5cm]{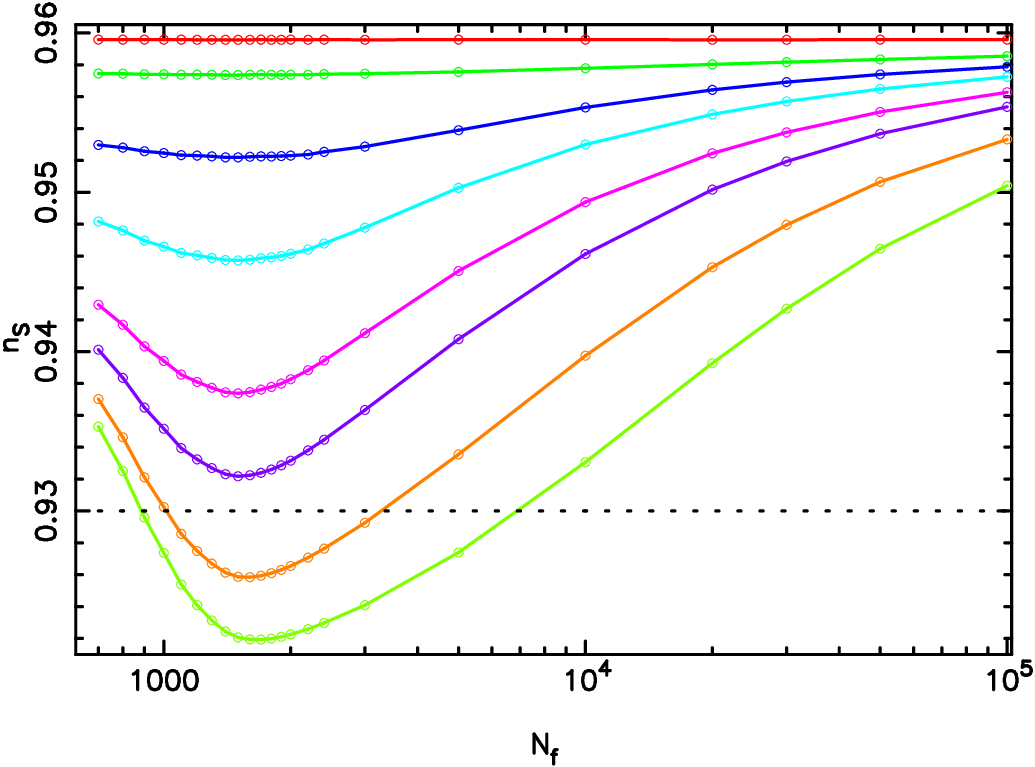}
\caption{\label{f:n} The spectral index $n_{\rm S}$ (from top to
bottom, $\beta$=0, 0.1, 0.3, 0.5, 0.7, 0.8, 0.9 and 0.95). The
dotted line shows the observational lower limit on $n_{\rm S}$ from
WMAP3.}
\end{figure}

The most important observable in Nflation models is the spectral
index, which we calculated as described above and show in
Fig.~\ref{f:n}. We see that it depends on both model parameters
$N_{\rm f}$ and $\beta$.

The curves have a generic shape where they first dip, and then, after
a minimum typically around $N_{\rm f} = 1500$, increase to join the
single-field value. The single-field value is always obtained in the
case $\beta=0$, in which all masses become identical and this result
is well known (e.g.~Ref.~\cite{LR}). These curves look a little
different from Fig.~4 in Ref.~\cite{KL1}, studying an exponential mass
spectrum where the values asymptoted to constants for large $N_{\rm
f}$. This is just because $N_{\rm f}$ is defined here in a rather
different way; in Ref.~\cite{KL1} increasing it added new fields at
the top of the mass range, and if they were heavy enough they fell to
their minima before observable scales left the horizon. The
distinction here is that when we increase $N_{\rm f}$ we are packing
more fields into the same mass interval, forcing us to the equal-mass
case.  That the curves begin to rise after $N_{\rm f} \simeq 1500$
indicates that fields heavier than the 1500th have typically reached
their vacuum state before the 50 $e$-foldings point.

To compare with observations, the analysis in Ref.~\cite{wmap3} shows
that for $r=8/50$, the 95\% confidence lower limit for $n_{\rm S}$
lies at about 0.93, shown as the dotted line in the figure.  Provided
$\beta \leq 0.8$, the spectral index is always large enough regardless
of the number of fields. For large $\beta$, the model is excluded only
for a range of $N_{\rm f}$ running from about one thousand to several
thousand. Accordingly, most of the model parameter space is currently
viable.

The results shown in Fig.~\ref{f:n} are the mean values over
realizations of the initial conditions. Additionally we find that the
spread in $n_{\rm S}$ values is quite small; the standard deviation of
$n_{\rm S}$ is never more than one percent of its displacement from
unity.  This confirms the existence of the thermodynamic regime for
this mass spectrum, just as in Ref.~\cite{KL1}.

\begin{figure}[t]
\centering
\includegraphics[width=7.5cm]{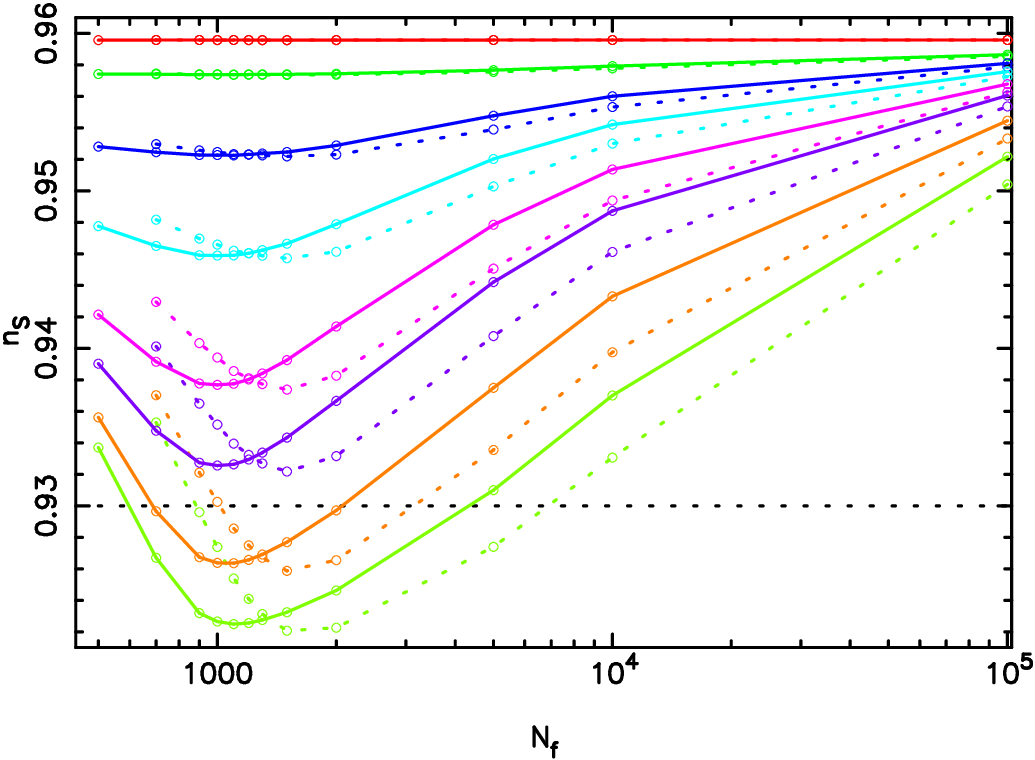}
\caption{\label{f:comp} As Fig.~\ref{f:n}, but now showing sets of
  curves for two choices of initial condition distributions. The
  dashed lines reproduce the curves from Fig.~\ref{f:n}, and the solid
  ones show initial conditions with $\phi_i^2/M_{{\rm Pl}}^2$ drawn
  from a uniform distribution.}
\end{figure}

Although the above analysis shows that for our chosen initial
distribution, the observational predictions are independent of
realization, one might further ask whether there is dependence on the
choice of that distribution. In order to test that, we carried out two
further series of simulations. In the first series, we took
$\phi_i^2/M_{{\rm Pl}}^2$ to be chosen uniformly between zero and
one. The results are shown in Fig.~\ref{f:comp}. The curves are
shifted to a smaller number of fields, retaining both their shape and
minima. We found that when the number of fields is multiplied by 3/2,
making the total number of $e$-foldings the same, these solid curves
become matched to the dotted ones. This can be related to the
different initial distributions by an approximate analytic argument
given in the Appendix.

\begin{figure}[t]
\centering
\includegraphics[width=7.5cm]{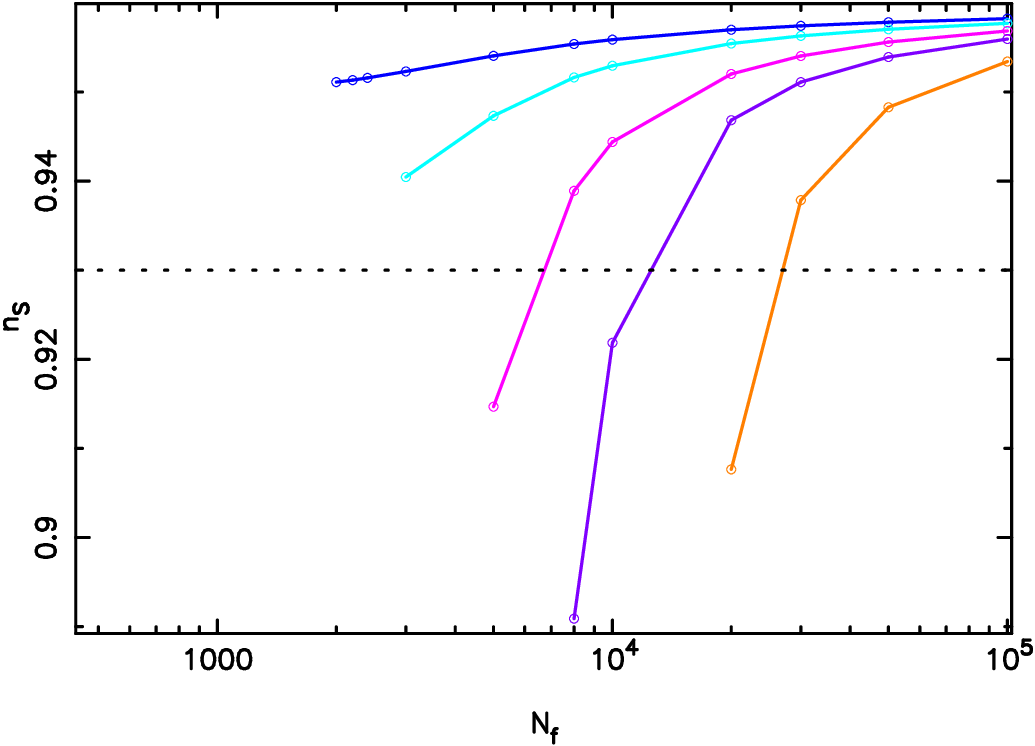}
\caption{\label{f:ne} As Fig.~\ref{f:n}, but showing results for
  uniformly-chosen field energy densities.  Only $\beta = 0.3$, $0.5$,
  $0.7$, $0.8$ and $0.9$ (from top to bottom) are shown. The left-hand
  end of the curves corresponds to models only just achieving 50
  $e$-foldings in total.}
\end{figure}

In the second series, we chose the field energy densities to be
uniformly distributed, with the lightest field ranging from $0$ to
$M_{{\rm Pl}}$ and the other fields given smaller ranges to ensure the
same expected energy density. This is distinct from Ref.~\cite{EM} who
gave each field the \emph{same} energy density, rather than the same
only on average.  In this case the more massive fields start closer to
their minima, and a greater total number of fields is needed to get
sufficient inflation.  The results are shown in Fig.~\ref{f:ne}. The
curves have a rather different shape, and there is some difficulty in
achieving a minimum of 50 $e$-foldings unless $N_{{\rm f}}$ is
large. The analytic analysis given in the Appendix does not apply in
this case, as the field initial values are correlated to the masses.
Nevertheless the overall conclusion is the same: that only for large
$\beta$ is there a danger that the models are ruled out through
predicting too small a value for $n_{{\rm S}}$, and even then only in
a narrow range of $N_{{\rm f}}$. The expected value $\beta \simeq 0.5$
is therefore viable for all initial condition distributions we tested.

\subsection{Other observables}

Concerning the tensor-to-scalar ratio $r$ and the non-gaussianity
parameter $f_{\rm NL}$, unsurprisingly the results are same as in our
previous work~\cite{KL1, KL2}. Comparing to the analytical results of
Eqs.~(\ref{e:r}), (\ref{e:fnl}), we confirm numerically that $r$ is
indeed independent of the model parameters, $r=0.16$, and that
$f_{\rm NL}\simeq 2/N$ which will never be sufficiently large to be
detected.  All simulations of $r$ and $f_{\rm NL}$ individually give
those results, without needing to average over realizations of the
initial conditions.

\section{Conclusions}

We have carried out a detailed numerical investigation of the
inflationary perturbations in Nflation models with the mass spectrum
of random matrix theory.  We produced the predictions for the spectral
index as a function of both the number of fields $N_{\rm f}$ and the
distribution parameter $\beta$ in the mass spectrum. We have found
that the model remains viable in the majority of its parameter space,
and that the thermodynamic regime, where the predictions become
independent of the initial condition realization, holds for this
spectrum as for the exponential case \cite{KL1}.


\begin{acknowledgments}
A.R.L.\ was supported in part by PPARC (UK). S.A.K.\ acknowledges the
hospitality of Filippo Vernizzi and the Abdus Salam ICTP in Italy, and
A.R.L.\ of the Institute for the Astronomy, University of Hawai'i,
while part of this work was being carried out.  We thank Richard
Easther for providing his code generating the MP mass spectrum, and
for many discussions relating to this work including the analytic
evolution formulation described in the Appendix.

\end{acknowledgments}


\appendix

\section*{Appendix: Analytic evolution}

This appendix is based on unpublished work by Richard Easther, whom we
thank for providing details.

In the main text we considered different initial distributions for the
scalar fields.  When the initial values of $\phi_i/M_{{\rm Pl}}$ and
$\phi_i^2/M_{{\rm Pl}}^2$ are randomly chosen uniformly between zero
and one, the expectation value $\alpha \equiv \langle \phi_i^2/M_{\rm
Pl}^2 \rangle $ is 1/3 and 1/2 respectively (other choices will give a
different $\alpha$).  Eq.~(\ref{e:efold}) becomes $N_{\rm tot}\simeq
N_{\rm f} \alpha/4$.  This difference by a factor 2/3 is closely
related to the shift in the curves shown in Fig.~\ref{f:comp}.

This can be shown via an analytic approximation to the evolution, in
order to extract estimates of the summations  $\sum_i^{N_{\rm f}}
m_i^2 \phi_i^2$ and $\sum_i^{N_{\rm f}} m_i^4 \phi_i^2$ which appear
in the expression for the spectral index. Using the slow-roll
approximation, the field value is
\begin{equation}
\phi_i (t) = \phi_i(t_0)\tau^{m_i^2/b}(t) \,,
\end{equation}
where $\tau(t)$ and $b$ have the same definition as in Ref.~\cite{EM}:
they are the ratio of the value of the heaviest field at the time $t$
to its initial value, and the upper limit of the probability
distribution of the mass-squared spectrum, respectively. Also $t_0$ is
the initial time. Then using the exponential function with $c=c(t)
\equiv 2 \ln [\tau(t)]/b$, we can rewrite the summation terms such as
\begin{equation}
\sum_i^{N_{\rm f}} m_i^2 \phi_i^2 (t) = \sum_i^{N_{\rm f}} m_i^2
\phi_i^2 (t_0)\exp [m_i^2 c]\,. 
\end{equation}
If one ignores correlations between the mass distribution and the
initial field distribution, so that $\langle m_i^2 \exp[m_i^2 c]
\phi_i^2(t_0) \rangle = \langle m_i^2 \exp[m_i^2 c] \rangle \, \langle
\phi_i^2(t_0) \rangle$, then with the Mar\u{c}enko--Pastur
distribution, the expectation value
is given by
\begin{equation}
\langle x^{i} \rangle = \bar{x}^i\hspace{1mm}{}_2F_1(1-i,-i;2;\beta)\,,
\end{equation}
where ${}_2F_1$ is the hypergeometric function. This follows from Eq.\
(6.14) of Ref.~\cite{EM}, where the summation can be rewritten as a
hypergeometric function.

Therefore the summation terms with the expectation values of field
initial conditions and of distribution of mass spectrum are
\begin{eqnarray}
&&\sum_i^{N_{\rm f}} m_i^2 \phi_i^2 (N) \label{e:m2phi2}\\
&&\hspace{5mm} = 4 N_{\rm tot} M_{\rm Pl}^2 \bar{m}^2\sum_{i=0}^{\infty}
  \bar{m}^{2i} \hspace{1mm}{}_2 F_1 (-i, -i-1;2;\beta)\frac{c^i}{i!}
  \,,\nonumber\\ 
&&\sum_i^{N_{\rm f}} m_i^4 \phi_i^2 (N) \label{e:m4phi2} \\
&&\hspace{5mm} = 4 N_{\rm tot} M_{\rm Pl}^2 \bar{m}^4\sum_{i=0}^{\infty}
  \bar{m}^{2i} \hspace{1mm}{}_2 F_1 (-i-1,
  -i-2;2;\beta)\frac{c^i}{i!}\,.\nonumber
\end{eqnarray}
Note that $t$ is related to $N_{\rm tot}-N$; that is, $t$ counts
forward from when the fields start to evolve from their own initial
conditions, while $N$ is the number of $e$-foldings before the end of
inflation.

The $n_{\rm S}$ data in Fig.~\ref{f:comp} are taken at different
times, but at the same $N$. Using Eqs.~(\ref{e:m2phi2}) and
(\ref{e:m4phi2}), the $n_{\rm S}$ formula becomes
\begin{equation}
n_{\rm S} \simeq 1 - \frac{1}{N} - \frac{f(t,\beta)}{N_{\rm tot}}\,, 
\label{e:n2}
\end{equation}
where
\begin{equation}
f(t,\beta) \equiv \frac{\sum_{i}^{\infty}\bar{m}^{2i} \hspace{1mm}
  {}_2 F_1 (-i-1, -i-2;2;\beta){c^i}/{i!}}{\left[\sum_{j}^{\infty}
  \bar{m}^{2j} \hspace{1mm}{}_2 F_1 (-j,
  -j-1;2;\beta){c^j}/{j!}\right]^2} 
\,. 
\end{equation}
Here the function $f(t,\beta)$ represents how $n_{\rm S}$ evolves with
respect to time (or $N$). Even though it is hard to evaluate it in
general, it will have the same value whenever $t$ and $\beta$ are
same.  Hence in cases with same $N_{\rm tot}$, but not necessarily the
same $N_{\rm f}$ or $\alpha$, then $n_{\rm S}$ must be same. This
explains why the solid curves in Fig.~\ref{f:comp} are matched with
dotted ones after multiplying 3/2 to $N_{\rm f}$, because this makes
$N_{\rm tot}$ the same. Different $\alpha$ shifts the $n_{\rm S}$
curves along the $N_{\rm f}$ axis, with bigger $\alpha$ shifting the
curves in the smaller $N_{\rm f}$ direction.

We can conclude then that, insofar as the approximations hold, $\beta
\lesssim 0.8$ should satisfy the current observations regardless of
the field initial condition distribution, while for larger $\beta$
some choices of $N_{\rm f}$ will be ruled out and that those $N_{\rm
f}$ values depend on the $\alpha$ value of the initial condition
distribution. However the above analysis relies on using a slow-roll
approximation for all fields, which is likely to become increasingly
inaccurate as increasing numbers of fields evolve towards their
minima, and assumes the mass and initial field value distributions to
be uncorrelated. The latter approximation fails badly for the uniform
field energy density initial conditions.


\end{document}